\documentclass[12pt,preprint]{aastex}
\usepackage{bbm}
\usepackage{mathrsfs}
\usepackage{amssymb,amsmath,float}
\usepackage{rotating}
\usepackage{color}

\newcommand\ha{\hat a}
\newcommand\had{\hat a^\dag}

\newcommand\hS{\hat S}
\newcommand\var{{\rm var}}
\newcommand\cov{{\rm cov}}

\newcommand\cT{\mathcal{T}}
\newcommand\cS{\mathcal{S}}
\newcommand\cSa{\hat{\mathcal{S}}^{(2)}}
\newcommand\cSb{\hat{\mathcal{S}}^{(3)}}
\newcommand\cSc{\hat{\mathcal{S}}^{(4)}}
\newcommand\cSn{\hat{\mathcal{S}}^{(n)}}
\newcommand\fjk{f_{jk}}
\newcommand\al{\alpha}

\newcommand\de{\delta}

\newcommand\ta{\tau}

\newcommand\om{\omega}

\newcommand\De{\Delta}

\newcommand\<{\langle}
\renewcommand\>{\rangle}

\newcommand\ie{\emph{i.e.}}
\newcommand\eg{\emph{e.g.}}

\newcommand\beq{\begin{equation}}
\newcommand\eeq{\end{equation}}
\newcommand\bea{\begin{eqnarray}}
\newcommand\eea{\end{eqnarray}}
\newcommand\bal{\begin{align}}
\newcommand\eal{\end{align}}

\newcommand\fr{\frac}

\newcommand\ap{\approx}

\renewcommand\bal{\mbox{\boldmath$\alpha$}}

\begin{document}

\title{Improvement in the accuracy of flux measurement of radio sources by exploiting an arithmetic pattern in photon bunching noise}

\author{Richard Lieu$^1$}

\affil{$^1$Department of Physics, University of Alabama,
Huntsville, AL 35899\\}

\begin{abstract}
A hierarchy of statistics of increasing sophistication and accuracy is proposed, to exploit an interesting and fundamental arithmetic structure in the photon bunching noise of incoherent light of large photon occupation number, with the purpose of suppressing the noise and rendering a more reliable and unbiased measurement of the light intensity.  The method does not require any new hardware, rather it operates at the software level, with the help of high precision computers, to reprocess the intensity time series of the incident light to create a new series with smaller bunching noise coherence length.  The ultimate accuracy improvement of this method of flux measurement is limited by the timing resolution of the detector and the photon occupation number of the beam (the higher the photon number the better the performance).  The principal application is accuracy improvement in the 
bolometric flux measurement of a
radio source.
\end{abstract}

\section{Introduction}

For observational astronomers to date, the principal carrier of information about the Universe is still electromagnetic waves.  Whether it be source or background radiation, it is usually collected (photometrically or spectroscopically) as a data stream of photon arrival times, readily convertible to a time series of the radiation intensity, the first two moments of which yield the source or background brightness and its variability.  For most applications, however, moments higher than the first two are not regarded as useful.  Indeed, in the case of steady emission, even the second moment itself seems to be redundant, as it can at best only provide an estimate of the measurement error of the mean intensity, which can in any event be calculated if the nature of the emission is known.  Thus, \eg~in optical and X-ray wavelengths the uncertainty in the mean intensity as given by the second moment is inferrable from the formula of Poisson statistics, while in the radio wavelength the radiometer equation.

Yet the logical steps taken above has another side to them which does not seem to have been fully exploited.  On one hand, one contends that the most important quantity being measured, the mean intensity of the arriving light, is strictly speaking the mean value taken over the finite time domain of an observation, {\it viz.} it is a {\it sample average} which even in the case of steady emission cannot be a more accurate representation of the true {\it ensemble} intensity unless the time series is replaced by one of longer duration, acquired by another observation involving more exposure time.  On the other hand, when estimating the ensemble uncertainty in the mean without involving the sample value of the next moment, one has tacitly accepted as axiomatic that the underlying emission is steady, and the pre-existence of this information affords one a {\it prior} model of the radiation, {\it viz.} that of stationary chaotic light.  In this case, it is the purpose of our paper to demonstrate, at least in the context of radio observations, that one could go a lot further by pursuing the argument in the opposite direction: the {\it ensemble} mean intensity can likewise be estimated from an exposure limited time series, in a manner more accurately than the sample mean alone, provided the process of inference involves the sample values of the higher moments.

The crucial point is that although for given statistical model the ensemble value of all the moments are computable from formulae pertaining to the model, the sample values can deviate independently of each other from their respective ensemble ones, as otherwise knowledge of a subset of the sample moments would be sufficient to enable one to deduce the entire time series, which is usually not the case.  Thus there {\it is} useful additional information in the higher moments; provided they are assimilated with the help of a {\it prior} model, it is possible to estimate the mean intensity of the source or background with much improved accuracy than conventional methods, at least in principle.

Unlike laser light, steady emissions of natural light from the sky is usually characterized by two basic properties: stationarity and incoherence (chaos).  The former means the time series of the light intensity exhibits no preferred instance of time, while the latter (interpreted in the sense of temporal coherence) means the same time series is the modulus square of many Fourier modes with different frequencies and random uncorrelated phases.  These basic properties are responsible for the existence of phase noise, also commonly referred to as photon bunching noise, {\it viz.} the intensity varies as $\de I/I \ap 1$ on timescales $\ta \ap 1/\Delta\om$, known as the coherence time, where $\Delta\om \ll \om_0$ is the frequency bandwidth of the light, which is either the intrinsic spectral width of the emitter or the filter response width of the instrument in the case of a broad band source.   Moreover, there is in general no intensity correlation among these segments of length $\ta$.  Owing to the particle aspect of light, there is another source of noise in addition to photon bunching, {\it viz.} shot noise or Poisson noise, which depicts the random arrival times of individual unresolvable pulses.  This noise is genuinely memoryless down to all scales however short, \ie~the time tag of one photon does not reflect at all how early or late the previous and next photon will arrive.   For reviews of these two fundamental noise sources in natural chaotic light sources, see \cite{boy82,lou00,zmu03}.

The relative importance of bunching noise and shot noise in astronomical sources is usually determined by the average number $n_0$ of photons collected by the telescope in the timescale of $\ta$.  If $n_0 \gg 1$, as in radio telescopes, bunching noise overwhelms shot noise, and the latter is insignificant.  For an observing time $\Delta t$ and filter bandwidth $\De\om$ satisfying $\Delta t \Delta\om \gg 1$, the accuracy of a bolometric flux measurement of a broad band emitter with spectrum covering a frequency range $> \De\om$ is then given by the number $N$ of segments of duration $\ta$ that spans the total exposure time $\Delta t$, specifically the ratio of the noise amplitude to the mean intensity goes as $1/\sqrt{N}$, which is part of the {\it radiometer equation} (\cite{bur10,chr85}) to be derived below as (\ref{errordirect}).  In the case of a narrow band source that radiates only in some restricted spectral range smaller than the filter limits, the radiometer equation would then apply with $\De\om$ representing the emission bandwidth itself.  On the other hand, at shorter wavelengths than the radio, $n_0$ quickly falls below the $n_0 =1$ limit, and it is the Poisson statistics of photon counting that limits the flux accuracy, \cite{bir06}.

In this paper we are primarily interested in the accuracy improvement of radio flux measurements (\ie~the $n_0 \gg 1$ limit of bunching noise domination), which could prove invaluable given the drastic limitation on the size of single dish telescopes that can still be built.  In recent literature the subject of photon bunching noise suppression was addressed in the context of radio flux measurements, see \cite{lie14,nai15,zmu15}.  The debate among these authors was whether the fundamental limit to the sensitivity of radio receivers given by the radiometer equation is surpassable.
As stated earlier, unlike shot noise which is a genuine memoryless and irreproducible quantum effect, bunching noise is determined {\it a priori} by a set of hidden variables, {\it viz.} the phase and amplitude of the constituent Fourier modes, and can be reproduced in \eg~a beam splitter, \cite{han57}.  This offers the tempting suggestion that the noise is susceptible to manipulation and control.
Hence, if there is any way of  reducing this noise in long wavelength observations, it would represent an improvement in the accuracy of flux determination of a radio source.

Complementary to and independently of the homodyne detection idea (which is currently being challenged by \cite{zmu15}), it is proposed in this paper a technique of analyzing the intensity time series data of incoherent radio signals, \ie~one which does not involve any extra hardware setup, rather operates at the software level, to take advantage of mathematical correlation effects embedded in the photon bunching noise in attempt to suppress it.  

Let us begin with a brief revisit to the theory of the two principal noise sources of incoherent light, bunching noise and shot noise.
So long as we are dealing with a narrow bandwidth, it is more convenient to work with the Fourier transforms of the annihilation and creation operators,
 \beq \ha(t) = \fr{1}{\sqrt{2\pi}} \int d\om\,\ha(\om)e^{-i\om t}, \qquad
 \had(t) = \fr{1}{\sqrt{2\pi}} \int d\om\,\had(\om)e^{i\om t}. \eeq
They satisfy the commutation relations
 \beq [\ha(t), \ha(t')] = \de(t-t'). \label{comm} \eeq
For a chaotic beam with Gaussian frequency profile, centred on $\om_0$ and with bandwidth $1/\ta$, we have
 \beq \<\had(t)\ha(t')\> = n_0 f(t-t'), \label{corr} \eeq
where $f(t)$ is given by
 \beq f(t) = \fr{1}{\ta} e^{i\om_0 t}e^{-t^2/2\ta^2} \label{fD} \eeq if the radiation spectrum has a strong and Doppler broadened line, or a broad band continuum which is narrowed down by a Gaussian spectral filter;  or
\beq f(t)= \fr{1}{\ta} e^{i\om_0 t}e^{-|t|/2\ta} \label{fC} \eeq if there is a collisionally broadened line, or a continuum selected by a Lorentzian filter, see section 3.4 of \cite{lou00}.
Moreover, the intensity is simply $\om_0 \had(t)\ha(t)$, but in the narrow-band case, it is simpler to remove the factor of $\om_0$, and talk instead about
 \beq \hS(t) = \had(t)\ha(t), \eeq
which represents the number of photons arriving per unit time.  It follows immediately that
 \beq \<\hS(t)\> = \fr{n_0}{\ta} \label{mean} \eeq is the mean photon rate.

Now we evaluate the covariance function
 \beq \cov(S(t), S(t')) = \<\hS(t) \hS(t')\> - \<\hS\>^2, \label{covJ} \eeq where
\bea
\<\hS(t) \hS(t')\> &=& \<\had(t)\ha(t)\had(t')\ha(t')\>\notag\\
&=& \<\had(t)\ha(t')\> \delta (t-t') + \<\had(t)\had(t')\ha(t)\ha(t')\> \notag\\
&=& \<\had(t)\ha(t')\> \delta (t-t') + \<\had(t)\ha(t')\>\<\had(t')\had(t)\> + \<\had(t)\ha(t)\>\<\had(t')\had(t')\> \notag\\
&=& \fr{n_0}{\ta}\de (t-t') + n_0^2 |f(t-t')|^2 + \fr{n_0^2}{\ta^2}.
\label{twopt}
\eea
The last but one step was taken assuming the noise of incoherent light obeys Gaussian moment theorem, as is the case when the light is maximally chaotic and the light curve is time homogeneous with no preferred $t$ (\ie~correlations depend only on time difference, not absolute time).  In general, this allows one to write \beq \<\had(t_1)\cdots\had(t_N)\ha(t'_1)\cdots\ha(t'_N)\> = \sum_P \<\had(t_1)\ha(t'_1)\>\cdots\<\had(t_N)
\ha(t'_N)\>, \label{gaussth} \eeq where $P$ stands for all possible $N\!$ pairings of with $t'_1, \cdots , t'_N$, see \cite{wan89}.

If we define the average flux over a short time interval as
 \beq \hS_T(t) = \fr{1}{T} \int_{t-T}^t dt'\,\hS(t'), \label{JT} \eeq
then we find
 \beq \var(S_T) =  \fr{1}{\ta T}\left[n_0^2 F\left(\fr{T}{\ta}\right) + n_0\right], \eeq
where
 \beq F\left(\fr{T}{\ta}\right) = \fr{\ta}{T} \int_{-T}^T dt\,(T-|t|)|f(t)|^2. \eeq
Note that if $T\ll \ta$, by (\ref{fD}) or (\ref{fC}) the function $f(t)$ in the integrand will be reduced to $1/\ta^2$, so $F(T/\ta)\ap T/\ta$.  The covariance between two measurements of $S$ which took place during intervals $T$ centered at times $t_j$ and $t_k$ is, from (\ref{covJ}) and (\ref{twopt}), \beq {\rm cov} (\hS_j,\hS_k) = \fr{n_0}{\ta} \de_{jk} + \fr{n_0^2}{\ta^2}\fjk,  \eeq or, ignoring the shot noise term $\de_{jk}$, \beq \fr{{\rm cov} (\hS_j,\hS_k)}{\<S_j\>^2} = \fjk, \label{fjk1} \eeq where the operator $\hS_j$ represents the measurement of the average photon arrival rate over the time interval $(j-1)T$ to $jT$, and \beq f_{jk}=e^{-(t_j-t_k)^2/\ta^2};~{\rm or}~e^{-|t_j-t_k|/\ta}, \label{fjk} \eeq \ie~depending upon whether $f(t)$ is given by (\ref{fD}) or (\ref{fC}) respectively.

The ratio of the variance in the measurement of $S_T$ to the square of the mean signal $\<\hS_T(t)\>^2$ is given by
 \beq \fr{\var(S_T(t))}{\<\hS_T(t)\>^2} = \fr{\ta}{T} \left[F\left(\fr{T}{\ta}\right)
 + \fr{1}{n_0}\right], \eeq
or
 \beq \fr{\var(S_T(t))}{\<\hS_T(t)\>^2} \ap 1+ \fr{\ta}{n_0 T},\quad\text{for}\quad T\ll\ta
  \label{errorJ}\eeq
which is $\ap 1$  if shot noise is ignored, in agreement with the $j=k$ limit of (\ref{fjk1}) as both indicates the error in a single short-term measurement is as large as the measurement.
On the other hand, if we measure for a much longer time $\cT=NT$, we may replace $T$ in (\ref{JT}) by $\cT$ and use the limiting value of $F(x)$ for $x\gg 1$, {\it viz.} $\sqrt{\pi}$ or $2$ if $f(t)$ has the form (\ref{fD}) or (\ref{fC}) respectively.
Then we have
 \beq \fr{\var(S_\cT(t))}{\<\hS_\cT(t)\>^2} \ap \sqrt{\pi} \fr{\ta}{\cT}=\sqrt{\pi} \fr{\ta}{NT} + \fr{\ta}{n_0 NT} \ap \sqrt{\pi} \fr{\ta}{NT},
 \quad\text{for}\quad \cT\gg\ta\gg T
 \label{errordirect}\eeq
and for $f(t)$ of the form (\ref{fD}); otherwise for $f(t)$ of the form (\ref{fC}) the factor of $\sqrt{\pi}$ is replaced by $2$.

In the high occupation number limit of \beq \fr{n_0 T}{\ta} \gg 1 \label{highocc} \eeq (despite $T \ll \ta$) the shot noise terms, \ie~ the last term of (\ref{errorJ}) and (\ref{errordirect}) may be ignored, and (\ref{errordirect}) is sometimes known as the radiometer equation, see \cite{bur10,chr85}.  It embodies the concept of coherence length of photon bunching noise, {\it viz.} the intensity time series of fully incoherent light (sometimes also known as Gaussian thermal light) comprises $N\ap\cT/\ta$ segments of approximately constant flux within, but randomly uncorrelated fluxes without.  Thus, while shot noise has zero coherence length and exhibits the $1/\sqrt{N'}$ behavior ($N' = n_0\cT/\ta$ is the mean number of photons arriving in time $\cT$) of the r.m.s. to mean ratio down to any scale, bunching noise has this behavior only on scales larger than the coherence length of $\ta$.  In the ensuing discussion we focus our attention upon incoherent light satisfying (\ref{highocc}).

\section{Bunching noise suppression: a relatively simple trial statistic}

Evidently, in (\ref{errordirect}) one sees that the shorter the coherence length $\ta$ for a given total exposure $\cT=NT\gg\ta$, the smaller the noise-to-signal ratio because there is more opportunity for intensity reversals among the various coherent segments to cancel out the noise.  But because $\ta$ is just the reciprocal of the frequency bandwidth of the radiation, there seems to be no way of changing it.  Here we investigate further if this conclusion is correct.

Let us consider an intensity time series $S_j$ comprising $N$ measurements lasting an interval $T\ll \ta$ each, such that the total data span is $\cT = NT \gg \tau$ as before, the mean intensity is given by (\ref{mean}), and the r.m.s. to mean ratio the square-root of (\ref{errordirect}).  But suppose one constructs the time series of \beq \cSa = \hS + \al \hS^2, \label{S2} \eeq where $\al$ is an adjustable parameter (note $\al$ is not dimensionless, and the reason for the superscript $(2)$ is the existence of an entire hierarchy of statistics $\cSn$ to be discussed in the next section, among which $\cSa$ is the first).  The mean of $\cSa$ is \beq \<\cSa\> = \<\hS_j\> + \al\<\hS_j^2\> = \fr{n_0}{\ta} + \fr{2\al n_0^2}{\ta^2}.  \label{m1} \eeq
For the covariance function of $\cSa$, we need to first calculate $\<\hS_j^2 \hS_k\>$, with the help of (\ref{comm}) and (\ref{gaussth}) and invoking the condition (\ref{highocc}), \beq \<\hS_j^2 \hS_k\> = \<\had_j\ha_j\had_j\ha_j\had_k\ha_k\> = 2\<\had_j\ha_j\>\<\had_j\ha_j\>\<\had_k\ha_k\>+4\<\had_j\ha_j\>\<\had_j\ha_k\>\<\had_k\ha_j\> = \fr{2n_0^3}{\ta^3} (1+2f_{jk}), \label{S2S} \eeq where $\fjk$ is as in (\ref{fjk}).

In general, the coefficient $\eta_p$ of the $f_{jk}^p$ term in $\<\hS_j^n \hS_k^m\>$, where $p \leq {\rm min} (n,m)$, equals the number of different ways of choosing $p$ annihilation operators $\ha$ out of the $n$ available ones from $\hS_j^n$ without paying attention to order, multiplied by a similar number of $p$ annihilation operators from $\hS_k^m$ (these two sets of operators may then switch sides) {\it and} the number of ways of re-ordering $\ha$ within $\hS_j^n$ and $\hS_k^m$, {\it viz.} the coefficient is given by the product of four quantities: \beq \eta_p = C^n_p C^m_p n! m! \left(\fr{n_0}{\ta}\right)^{m+n}, \label{coeff} \eeq where the last is factor is merely a normalization factor.

With the help of (\ref{coeff}) one may evaluate $\<\hS_j^2 \hS_k^2\>$, as  \beq \<\hS_j^2 \hS_k^2\> = \fr{n_0^4}{\ta^4}(4+16f_{jk}+4f_{jk}^2). \label{S2S2} \eeq  It is now possible to assemble all relevant terms to derive the covariance function as \beq {\rm cov} (\cSa_j,\cSa_k) = \fr{n_0^2}{\ta^2}f_{jk}\left(1+\fr{4n_0}{\ta}\al\right)^2 + \fr{4n_0^4}{\ta^4}\al^2 f_{jk}^2.  \label{cov1} \eeq   From (\ref{m1}) and (\ref{cov1}), one sees that by setting \beq \al = -\fr{1}{4\<\hS_j\>} = -\fr{\ta}{4n_0}, \label{alpha1} \eeq the mean of $\cSa_j$ becomes $\<\cSa_j\> = n_0/(2\ta) = \<\hS_j\>/2$ and the covariance is given by last term of (\ref{S2S2}).  In other words, one obtains the relatively simple result \beq \fr{{\rm cov} (\cSa_j,\cSa_k)}{\<\cSa_j\>^2} = \fjk^2, \label{fjk2} \eeq which indicates the time series of $\cSa$ has a narrower autocorrelation function than $\hS_j$, since the latter is given by the first power of $\fjk$ (see (\ref{fjk1})).  Thus, even though (\ref{fjk2}) for $j=k$ yields $1$ as the ratio of the r.m.s. to the mean signal which is no improvement from (\ref{fjk1}), the {\it average} over many data points $j$ would have a smaller ratio of the same.

To elaborate, for the choice of $\al$ given by (\ref{alpha1}) the statistic $\<\cSa_j\>$ is an unbiased estimator of the mean intensity $\<\hS_j\>$ with normalized noise variance given by \beq \fr{{\rm var} (\cSa_\cT)}{\<\cSa_j\>^2} = \fr{4\ta^2}{N^2 n_0^2}\sum_{k,l=1}^{N} {\rm cov} (\cSa_k,\cSa_l) = \sqrt{\fr{\pi}{2}}\fr{\ta}{NT}~{\rm or}~\fr{\ta}{NT}, \label{var1} \eeq for a time series spanning $\cT = NT \gg \ta$, where the continuum representation \beq \sum_j e^{-2j^2T^2/\ta^2} \ap \fr{1}{T}\int dt\,e^{-2t^2/\ta^2} = \sqrt{\fr{\pi}{2}}\fr{\ta}{T} \label{cont} \eeq was employed, and the two possibilities on the right side of (\ref{var1}) correspond to the scenarios of $f(t)$ having the form (\ref{fD}) or (\ref{fC}) respectively.  Comparing (\ref{var1}) to (\ref{errordirect}), $\cSa_j$ appears to be a more accurate estimator of the mean intensity than $\hS_j$ itself, and this is because when (\ref{alpha1}) holds the only contribution to the covariance (\ref{cov1}) is from the $f_{jk}^2$ term, which is a {\it narrower} correlation function than $f_{jk}$.  In other words, the coherence length of $\cSa_j$ is less than $\hS_j$ with the appropriate choice of $\al$.

Yet there is a caveat: (\ref{m1}) yielded a time series of less coherence length and a mean of half the original time series, {\it viz.} $\<\cSa_j\> = \<\hS_j\>/2$, only if the value of $\al$ in it is set to (\ref{alpha1}), which presupposes {\it a priori} knowledge of $\<\hS_j\>$, the very quantity one is trying to measure.  In practice, one could always use the sample average of the $\hS_j$ series as starting point, and tune $\al$ to align the flux inferred from (\ref{alpha1}) with the sample average of the ensuing $\cSa_j$ series.  The error in the estimate would then be given by (\ref{var1}), which is an improvement over enlisting $\hS_j$ alone.

Next, we proceed to show
that there exists a whole hierarchy of statistics $\cSn$, with $\hat{\cal S}^{(1)} = \hS_j$ being the first, capable of delivering increasingly accurate estimates of $\<S_j\>$,
eventually surpassing (\ref{errordirect}).  This claim will be demonstrated in the rest of the paper.

Before doing that, a cautionary note is in order.  The proposed statistic $\cSa$ should work as described, if the incoherent light in question has high occupation number $n_0$, not simply in terms of $n_0 \gg 1$ but rather (\ref{highocc}).  To elaborate, note the validity of (\ref{cov1}) is contingent upon (\ref{highocc}), as (\ref{cov1}) ignores higher order terms, terms which depict extra, more complicated effects of both bunching noise and shot noise. In fact, the next lowest order term beyond the $O(n_0^2/\ta^2)$ we used is $O(n_0/(T\ta))$ which, if included, would modify (\ref{cov1}) to \beq {\rm cov} (\cSa_j,\cSa_k) = \fr{n_0^2}{\ta^2}f_{jk}\left(1+\fr{4n_0}{\ta}\al\right)^2 + \fr{4n_0^2}{\ta^4}\al^2 f_{jk}^2 - \fr{2\al n_0^2}{T\ta^2} (f_{jk} +4\de_{jk}) + \fr{8\al^2 n_0^3}{T\ta^3} (f_{jk}+3\de_{jk}) +\fr{n_0}{\ta}\de_{jk}, \label{cov2} \eeq it reduces to \beq {\rm cov} (\cSa_j,\cSa_k) = \fr{n_0^2}{4\ta^2}f_{jk}^2 + \fr{n_0}{2T\ta}\de_{jk} \label{cov3} \eeq when $\al$ is given by (\ref{alpha1}).  Thus, to justify dropping the next order correction to (\ref{cov1}) as given by the last term, (\ref{highocc}) has to be enforced to ensure the interval $T$ is too short.  This point will become particularly relevant when we proceed to investigate much more powerful statistics that stem from the ideas presented above.

\section{A hierarchy of statistics of increasing accuracy}

Apart from $\cSa$, there are even more sophisticated statistics belonging to the same family.  The immediate follow-up is  \beq \cSb_j = \hS_j + \al\hS_j^2 + \beta\hS_j^3. \label{S3} \eeq Here the calculation of the mean and covariance of $\cSb$ is more tedious, but requires no fundamental algebraic relations beyond (\ref{comm}) and (\ref{corr}), the manipulations of which were given in previous examples like (\ref{twopt}) and (\ref{S2S}).  Thus we will not be showing such intermediate steps again; we simply quote the results: \beq \<\cSb_j\> = \fr{n_0}{\ta}(1+2\al'+6\beta'), \label{cS3m} \eeq and
\bea {\rm cov} (\cSb_j,\cSb_k) &=& [(1+8\al'+16\al'^2+36\beta'+324\beta^{'2}+144\al'\beta')\fjk+\notag\\
& &(4\al^{'2}+72\al'\beta'+324\beta^{'2})\fjk^2+
36\beta^{'2}\fjk^3], \label{cov4} \eea where $\al'= \al n_0/\ta = \al\<\hS_j\>$, $\beta'=\beta n_0^2/\ta^2=
\beta\<\hS_j\>^2$ are dimensionless versions of $\al$ and $\beta$, and $\fjk$ is as in (\ref{fjk}).

There is again an embedded arithmetic structure that affords one a unique solution for $\al$ and $\beta$, corresponding to \beq \al' = \al s' = -\tfrac{1}{2},~{\rm and}~\beta' = \beta s'^2 =\tfrac{1}{18} \label{ab} \eeq such that when $s'=s$, where $s=\<\hS_j\>$ is the ensemble flux, $\<\cSb\> = n_0/(3\ta)$ and \beq \fr{{\rm cov} (\cSb_j,\cSb_k)}{\<\cSb_j\>^2} = \fjk^3. \label{fjk3} \eeq Comparing to (\ref{fjk1}) and (\ref{fjk2}), the autocorrelation of $\cSb$ contains only the $\fjk^3$ contribution with narrower width, or shorter coherence length, than $\cS^{(1)}_j = \hS_j$ by the factor of $\sqrt{3}$ respectively, assuming $\fjk$ is given by (\ref{fD}), if it is given by (\ref{fC}) then the factor will be $3$.  As before also, what actually happens in practice is that one achieves this next level of error reduction by using the flux estimated from the $\hS_j$ and $\cSa_j$ series as starting point, and tune $\al$ and $\beta$ via $s'$ to align $s'$ with the flux value inferred from the sample average of the ensuing $\cSb_j$ series.
The long time average of the $\cSb$ series would therefore lead to \beq \fr{{\rm var} (\cSb_\cT)}{\<\cSb_j\>^2} = \fr{9\ta^2}{N^2 n_0^2}\sum_{k,l=1}^{N} {\rm cov} (\cSb_k,\cSb_l) = \sqrt{\fr{\pi}{3}}\fr{\ta}{NT}~{\rm or}~\fr{2}{3}\fr{\ta}{NT}, \label{var2} \eeq corresponding to the scenarios of $f(t)$ having the form (\ref{fD}) or (\ref{fC}) respectively, which is a progressive improvement from (\ref{var1}).
Just like the statistic $\cSa_j$, (\ref{fjk3}) holds only when the incoherent light satisfies the high occupation number condition of (\ref{highocc}).


The pattern emerging from $\cSa$ and $\cSb$ persists onto $\cSc_j = \hS_j + \al\hS_j^2 + \beta\hS_j^3 + \gamma\hS_j^4$.  In this instance, a choice of the dimensionless coefficients $\al' = -3/4$, $\beta' = 1/6$, and $\gamma' = -1/96$ would yield
\beq \<\cSc_j\> = \fr{n_0}{4\ta};~\fr{{\rm cov} (\cSc_j,\cSc_k)}{\<\cSc_j\>^2} = \fjk^4, \label{fjk4} \eeq  and hence \beq \fr{{\rm var} (\cSc_\cT)}{\<\cSc_j\>^2} = \fr{16\ta^2}{N^2 n_0^2}\sum_{k,l=1}^{N} {\rm cov} (\cSc_k,\cSc_l) = \sqrt{\fr{\pi}{4}}\fr{\ta}{NT}~{\rm or}~\fr{1}{2}\fr{\ta}{NT}, \label{var3} \eeq corresponding to the scenarios of $f(t)$ having the form (\ref{fD}) or (\ref{fC}) respectively, which is an improvement over (\ref{var2}).


It is now possible to prove that an underlying order exists, {\it viz.} for all $n\geq 1$ there exists a unique flux measurement operator \beq \cSn_j = \hS_j +\al_1\hS_j^2 +\cdots +\al_n\hS_j^n \label{cSn} \eeq with \beq \<\cSn\> = \fr{n_0}{n\ta};~\fr{{\rm cov} (\cSn_j,\cSn_k)}{\<\cSn_j\>^2} = \fjk^n \label{fjkn} \eeq which as $n\to\infty$ creates from the original time series one with arbitrarily short coherence length for the bunching noise.
To establish formally there indeed exists an unlimited hierarchy of arbitrarily large $n$, however, it is necessary to find the general formula of the normalized coefficients \beq \al'_l =\al_l \<\hS_j\>^{l-1},~l=1,2,\cdots,n, \label{alnorm} \eeq that removes at $s' =s$ all the $\fjk^p$ terms in ${\rm cov} (\cSn_j,\cSn_k)$ except $\fjk^n$.  To this end, it is necessary to enlist another formula
\beq \<\hS_j^m \hS_k^n\> = \left(\fr{n_0}{\ta}\right)^{m+n}\sum_{p=0}^{{\rm min}(m,n)} C^n_p C^m_p n! m! \fjk^p \label{SjmSkn} \eeq which follows simply from (\ref{coeff}) and the fact that each $\fjk^p$ coefficient in ${\rm cov} (\cSn_j,\cSn_k)$ is the product of the corresponding one in (\ref{SjmSkn}) and the quantity $\al'_k \al'_m$, {\it viz.} \beq {\rm coefficient~of}~\fjk^p~{\rm term} = \sum_{r=p}^n \sum_{s=p}^n (-1)^{r+s} C^r_p C^s_p r! s! \al'_r \al'_s. \label{fjkcoeff} \eeq  This leads to \beq \al'_l = -(-1)^l \fr{C^n_l}{nl!},~l=1,2,\cdots,n \label{alphal} \eeq as a solution that retains only the $\fjk^n$ term in ${\rm cov} (\cSn_j,\cSn_k)$.  To elaborate how one arrives at (\ref{alphal}),  one may expand the expression \beq [1+x(1+y)]^n [1+x(1+z)]^n = \sum_{r=0}^n \sum_{s=0}^n \sum_{p=0}^r \sum_{q=0}^s  C^n_r C^n_s C^r_p C^s_q x^{r+s} y^p z^q \label{polynomial} \eeq with $C^r_p = 0$ for $r<p$; then set $x=-1$ so that only the $y^p y^q$ term with $p=q=n$ survives.  Since the coefficients of the $p=q<n$ terms are \beq \sum_{r=p}^n \sum_{s=p}^n (-1)^{r+s} C^n_r C^n_s C^r_p C^s_p = 0,~p=1,2,\cdots,n-1, \label{soln} \eeq and these coefficients must vanish, one deduces that (\ref{alphal}) is a solution (the overall normalization factor $1/n$ is there to ensure the correct coefficient of the only surviving term $\fjk^n$ in ${\rm cov} (\cSn_j,\cSn_k)$; indeed one can verify that (\ref{alphal}) reproduces \eg~the $\al'$, $\beta'$, and $\gamma'$ values given in the earlier examples of $\cSn$).

Proceeding next to prove the validity of (\ref{fjk2}), (\ref{fjk3}), (\ref{fjk4}) and so on without limit, one must calculate the mean signal $\<\cSn_j\>$ with $\alpha'_l$ given by (\ref{alphal}), and $\cSn_j$ by (\ref{cSn}) with $\al$ by (\ref{alnorm}).  Since $\<S_j^r \> = r!n_0^r/\ta^r$, this is the same as showing \beq \sum_{r=1}^n (-1)^r C^n_r =  -1, \label{iden} \eeq a result that follows readily from the expansion of $(1+x)^n$ with $x=-1$.  In this way a universal method of flux estimation, applicable to any high order $n$, becomes apparent: the minimum of the statistic $s^{' 2n-2} {\rm var} (\cSn)$ will, as $n \to\infty$, locate the true flux $s'=s$ to within an error defined by the equation \beq \fr{1}{n^2}~\sum_{p=1}^{n-1} (C^n_p)^2 \left(\fr{s'-s}{s}\right)^{2p} = \fr{1}{n^2}.  \eeq Equivalently for $n \gg 1$, the equation is \beq [I_0 (z) -1] =1, \label{I0} \eeq where $z=2n(s'-s)/s$ and $I_0$ is a modified Bessel function of the first kind.  In fact, the solution to (\ref{I0}) is $z \ap 1.81$, leading to \beq \Big|\fr{s'-s}{s}\Big| = \fr{0.905}{n}, \label{largen} \eeq which, for sufficiently large $n$, {\it surpass} (\ref{errordirect}) when
\beq
n^2 \gtrsim
\fr{NT}{\ta}. \label{crit1} \eeq  This means, in other words, (\ref{largen}) is the most conservative flux sensitivity limit for large $n$, even though it turns out the same equation also works for $n=2,3,\cdots$ with negligible loss of accuracy.  In practice, however, as can be seen from (\ref{var1}), (\ref{var2}) and (\ref{var3}), the actual method of locating $s$ need not be as elaborate as minimizing $s^{' 2n-2} {\rm var} (\cSn)$, rather simply tuning $\al_l$ via $s'$ to align $s'$ with the value of the flux inferred from the ensuing $\cSn_j$, and may lead to even smaller flux uncertainties than the (\ref{largen}) limit.

\section{Upper bound on the accuracy improvement}



\begin{figure}
\begin{center}
\includegraphics[angle=0,width=6.5in]{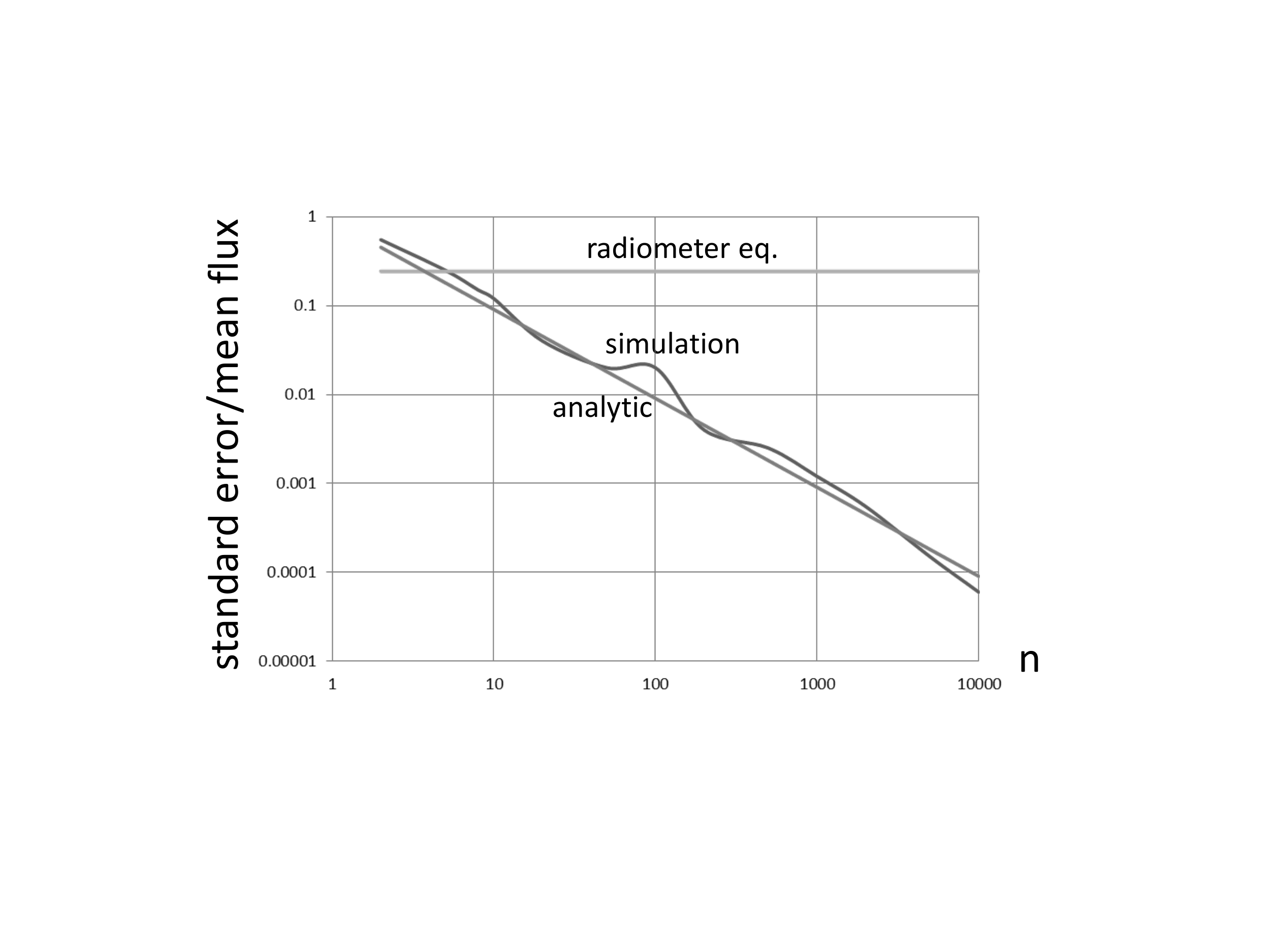}
\end{center}
\caption{Accuracy curves of the mean intensity of a time series of stationary chaotic light in radio frequencies as inferred by two methods: the radiometer equation of (\ref{errordirect})
(grey line) and the new method of this paper.
The coherence length of the time series is assumed to be $\ta = 1$~ns (bandwidth $\ap 1$~GHz); and the sampling time $T = 1$~ps (Doppler/Gaussian light) or $T=10$~fs (collisional/Lorentzian light). The total duration of the time series is $NT = 30$~ns, \ie~the number of samples is $N=3\times 10^4$ (Doppler/Gaussian light) or $N=3\times 10^6$ (collisional/Lorentzian light).  The wavy line is from a simulation of the Doppler/Gaussian light under the same conditions as stated above (for further details see text).}

\label{cartoon}
\end{figure}

From the foregoing discussion it would seem, provided one is prepared to work on the hierarchy of statistics $\cSn$ with an arbitrarily large $n$, the bunching noise error (\ref{largen}) in one's flux estimate can be reduced to a level far beneath the radiometer equation (\ref{errordirect}) governing the original time series $\hS_j$.

In reality there is a limit to how large $n$ can be, however, as one's ability in pursuing the new method
depends crucially upon resolving the bunching variations at the minimized coherence length in the ensuing $\cSn_j$ series, {\it viz.} length $\ap \ta/\sqrt{n}$ for Doppler/Gaussian light, (\ref{fD}), or $\ap \ta/n$ for collisional/Lorentzian light (\ref{fC}).  This requires a sampling interval $T$ for $\hS_j$ small enough to satisfy \beq T \ll \fr{\ta}{\sqrt{n}}~{\rm (Doppler/Gaussian)}~{\rm or}~T \ll \fr{\ta}{n}~{\rm (collisional/Lorentzian)}. \label{crit2} \eeq  Combining with (\ref{crit1}), it also means \beq N\ll \fr{\ta^5}{T^5}~{\rm (Doppler/Gaussian)}~{\rm or}~N\ll \fr{\ta^3}{T^3}~{\rm (collisional/Lorentzian)}. \label{crit2} \eeq
Yet, this need for ever higher timing resolution in the individual flux measurements of $\hS_j$ would eventually become difficult to meet technologically; thus if $\ta \ap 1$~ns as is typical of radio astronomy, $n \ap 10^6$ would demand femtosecond sampling.  In any case, incoherent detectors (radiometers) with the required combination of speed and sensitivity do not currently exist and would be challenging to design and build.
Moreover, unless the incident beam has infinite occupation number, criterion (\ref{highocc}) would also break down when $T$ is too small, and such extra contributions to the intensity noise as terms in (\ref{cov2}) complicate the performance of the method.

Overall, for radio wavelengths $n \ap 10^6$ is likely to be the absolute maximum one could envisage.  But if $n$ is much smaller, to the point (\ref{crit1}) no longer holds, the proposed new method will have no advantage over the radiometer equation.  The way to overcome this is to divide the time series into a number of segments\footnote{The other advantage of analyzing small segments at a time is that it minimizes the need to store large quantities of data at any given time.},
each having sufficiently small $N$ to satisfy (\ref{crit1}), then apply the new method to get the mean flux for each segment before taking the global average.  An improvement in flux accuracy by 10 -- 100 times, as illustrated by the scenario of Figure 1, should quite readily be achievable by sampling and digitizing each time stream (segment) at the rate of $10^{-3}$ to $10^{-4}$ of a coherence time.  Note that in Figure 1 we also included the flux determination accuracy of a numerical simulation of Doppler/Gaussian light with uncorrelated phases among the modes and autocorrelation function given by (\ref{fD}),  under the same conditions as stated in the caption.  The simulation comprises 30 independent realizations of the intensity time series, and in each case the fluxes as estimated by the radiometer equation and the new method are compared to the true flux of the simulation model to determine the accuracy of each method.  The ratio of the r.m.s. deviation between the true (ensemble) flux and the flux estimated by each method is plotted as the $y$-coordinate.


Lastly, we wish to point out that the conclusion of this paper does not necessarily contradict \cite{nai15}, who demonstrated the impossibility of surpassing the sensitivity limit of the radiometer equation (\ref{errordirect}) by the use of any unbiased estimator of the flux, because
the method we proposed here is more sophisticated than just employing a flux estimator, rather it tunes the flux to minimize a very high order statistic.  Thus the two approaches are fundamentally different.

\end{document}